\documentclass{aastex}
\usepackage{emulateapj5}

\slugcomment{To appear in ApJ Letters}

\shorttitle{UV and mid-IR emission in high-z IR galaxies}
\shortauthors{Charmandaris, Le Floc'h, \& Mirabel}

\begin{document}

\title{A bias in optical observations of high redshift luminous infrared galaxies}

\author{V. Charmandaris\altaffilmark{1}}
\affil{Astronomy Department, Cornell University, Ithaca, NY 14853, USA}
\email{vassilis@astro.cornell.edu}

\author{E. Le Floc'h\altaffilmark{2}, I.F. Mirabel\altaffilmark{3}}
\affil{CEA/DSM/DAPNIA, Service d'Astrophysique, F-91191 Gif-sur-Yvette, France}
\email{elefloch@as.arizona.edu,fmirabel@cea.fr}


\altaffiltext{1}{Chercheur Associ\'e, Observatoire de Paris, LERMA, 61 Av. de l'Observatoire, F-75014 Paris, France}

\altaffiltext{2}{Now at Steward Observatory, University of Arizona, 933 North Cherry Avenue, Tucson, AZ 85721, USA}

\altaffiltext{3}{Instituto de Astronom\'\i a y F\'\i sica del Espacio, cc 67, suc 28. 1428 Buenos Aires, Argentina}

\begin{abstract}
We present evidence for the dramatically different morphology between
the rest frame UV and 7$\mu$m mid-IR emission of VV\,114 and Arp\,299,
two nearby (z$\sim$0) violently interacting infrared luminous galaxies
(LIRGs). Nearly all LIRGs are interacting systems and it is currently
accepted that they dominate the IR emission at z$>$1. Luminous IR
galaxies located at z=1--2 could easily be detected as unresolved
sources in deep optical/near-IR ground based surveys, as well as in
upcoming 24$\mu$m surveys with the Space Infrared Telescope
Facility. We demonstrate that the spatial resolution of these surveys
will result in blending of the emission from unresolved interacting
components. An increased scatter will thus be introduced in the
observed optical to mid-IR colors of these galaxies, leading to a
systematic underestimation of their dust content.

\end{abstract}


\section{Introduction}

Over the past decade several studies on the physical characteristics
of high redshift galaxies have revealed a number of new exciting
results. In particular it has been established that the star formation
rate (SFR) of both individual galaxies, as well as the SFR per
co-moving volume, increases by more than an order of magnitude as we
move from systems of the local Universe to those found to a redshift
of z$\ge$2 \citep[e.g.,][]{Steidel96,Madau98}. The details on exactly
how the SFR evolves at even higher redshifts are still under debate as
they are strongly affected by small number statistics on the
properties of the population of the sources detected in sub-mm
surveys, but more importantly by the accurate modeling of the effects
of dust extinction. A number of deep surveys --- most notably the
Hubble Deep Field --- have also shown that a large number of faint
galaxies at intermediate and high redshift display irregular and/or
disturbed morphology as well as an excess of blue light emission
\citep[see][]{Williams96,Abraham96,LeFevre00}. A way to explain this
faint blue excess in the galaxy luminosity function is by introducing
a rapidly evolving population of star forming galaxies at those
redshifts \citep{Glazebrook95,Metcalfe01}. What are the properties
though, of this population? Several researchers suggest that these
distant galaxies are a new class of systems which are physically small
in size \citep{Pascarelle96}, while others speculate that they are
high redshift analogues of local interacting galaxies. The latter
scenario can be understood since those distant galaxies may look
peculiar because of the k-correction and band shifting of their short
wavelength (UV) emission to the optical HST filters \citep[][~and
references therein]{Hibbard97}.

Two additional independent findings seem to suggest that galaxy
interactions may indeed have significant implications to the blue
excess in the galaxy luminosity function. The first one is the ample
evidence that the galaxy interaction/merge rate increases as a power
law function of redshift $\sim$(1+z)$^m$
\citep{Carlberg90,Lavery96}. Even though a direct observational
measure of the interacting/merging galaxy pairs, and hence $m$, has
been difficult due to a number of potential biases in surveys, current
estimates indicate that $m$ can be as high as 3--4.5 \citep[see][~for
a review]{Lavery96,LeFevre00}. The second is that phenomena associated
with galaxy interactions such as the presence of starburst, infrared
luminous, and ultraluminous (LIRGs/ULIRGs) galaxies detected with IRAS
also increase with
redshift\citep[e.g.,][]{Lonsdale90,Sanders96,Xu01}. Tidally triggered
massive star formation in galaxies, which may eventually transform gas
rich galaxies into their ultraluminous phase, will be responsible for
increasing their rest frame UV flux and hence for the blue color
excess of the galaxies detected in deep surveys. It is also the reason
why large quantities ($\sim$10$^9$M$_{\sun}$) of molecular gas may be
compressed over an area of few hundred pc around their nuclei
\citep{Downes98,Bryant99} eventually obscuring them behind large
quantities of dust. As a result, wavelength dependent extinction by
dust grains distorts the morphology of interacting galaxies, limiting
our ability to accurately probe their properties .  A prime example of
this phenomenon was displayed in the study of NGC\,4038/39 (``The
Antennae'') by \citet{Mirabel98}. This system of interacting galaxies,
located at distance of 20\,Mpc is only moderately infrared
luminous\footnote{We use the standard definition of L$_{\rm
IR}$(8--1000\,$\mu$m) = 5.62$\times$10$^5$\,D(Mpc)$^2$\,(13.48{\it
f}$_{12}$+5.16{\it f}$_{25}$+2.58{\it f}$_{60}$+{\it
f}$_{100}$)\,L$_\sun$ \citep[see][]{Sanders96},
H$_0$\,=\,75\,km\,s$^{-1}$\,Mpc$^{-1}$, and q$_o$\,=\,0.5 to
facilitate comparison with earlier work.}  (L$_{\rm
IR}$=4.9$\times10^{10}$\,L$_{\sun}$) and still contains considerable
quantities of molecular gas dispersed over a large area, which may
become available for a more intense star formation event in the future
\citep{Wilson00}. Despite that, half of the mid-IR flux of the system
originates from the optically obscured overlap region between the two
nuclei \citep[see][]{Mirabel98}.

It is thus reasonable to pose the following questions. How would the
morphology of systems more IR luminous than NGC\,4038/39 be affected
by the presence of dust and what consequences this may have in
measurements of the {\em spatially unresolved} IR luminous galaxies
which are readily discovered at high (z$\gtrsim1.5$) redshift? More
specifically, how might this affect the conclusions of color
correlations between a number of optical and near-IR surveys that are
performed from the ground \citep[e.g.,][]{Jannuzi99}, and those
scheduled to be performed with the Space Infrared Telescope Facility
(SIRTF) \citep[e.g.,][]{Rieke00,Lonsdale03}?

We will base our analysis on high resolution UV and mid-IR
observations of Arp\,299 and VV\,114 that became recently
available. Both galaxies have IR luminosities almost an order of
magnitude higher than NGC\,4038/39 (L$_{\rm
IR}\sim$5$\times10^{11}$L$_\sun$ for Arp\,299 and
$\sim$4$\times10^{11}$L$_\sun$ for VV\,114) but the projected spatial
separation between their two interacting components is still
sufficiently large (8 and 6\,kpc respectively) for us to be able to
resolve them given their proximity (41 and 80\,Mpc respectively).

\section{Observations}

All observations used in this paper were retrieved from previously
published work and we refer the reader to those publications for
details on the data reduction. The mid-IR data were obtained with
ISOCAM on board the Infrared Space Observatory (ISO) and were
presented by \citet{LeFloch02} for VV\,114 and by
\citet{Charmandaris02} for Arp\,299. The UV data of both galaxies were
retrieved from the HST archive. An analysis of the STIS images of
VV\,114 and FOC images Arp\,299 is included in \citet{Goldader02} and
\citet{Meurer95} respectively.

\section{Discussion}

\subsection{Dust and interactions at high-z}

A detailed study of the apparent morphology of peculiar galaxies as a
function of redshift was already performed by \citet{Hibbard97} by
examining UV and optical images of a sample of 4 interacting
galaxies. These authors found that at z$>$1.5 it is no longer possible
to discern features such as long tidal tails and bridges that would
identify these systems as clearly ``interacting'', even in exposures
as deep as the Hubble Deep Field. These authors also alluded that the
redshifted rest UV light, which will be sampled in the optical if
those galaxies are at z$>$1.5, can not be used to accurately image the
evolved stellar population.

The limitations discussed by \citet{Hibbard97} become even stronger if
one considers dust extinction\footnote{We note that if we were to
trace dust using the usual extinction law A$_V$, then A$_{\rm
UV}$$\sim$7A$_{\rm H_\alpha}$$\sim$270A$_{7\mu m}$
\citep{Mathis90}.}. Even though numerous z$\sim$2--3 galaxies have
been detected as UV drop outs, the actual magnitude of extinction
effects in the detection of distant galaxies using their rest frame UV
emission is still under debate \citep[see][~for conflicting
views]{Meurer99,Adelberger00}. However, a number of recent results
clearly suggest the presence of considerable quantities of dust at
high redshifts. Based on a UV selected sample of 16 galaxies at
2$\leq$z$\leq$2.6, \citet{Erb03} found that the SFR, measured using
the redshifted into the K-band H$\alpha$ line, is not only elevated
(16\,M$_{\sun}$yr$^{-1}$) but it is also a factor of $\sim$2.4 higher
than that estimated using their UV continuum (SFR$_{\rm UV}$). This
discrepancy in estimating the SFR is even higher for the systems with
low UV luminosity and can be easily understood due to the presence of
dust. Unfortunately no H$\beta$ line was detected from these galaxies
to measure their Balmer decrement, and consequently we can not be
certain whether the factor of 2.4 is just a lower limit. Furthermore,
UV selected samples may be biased towards systems with a low dust
content. A similar underestimate in SFR$_{\rm UV}$ compared to the one
calculated using the H$\beta$ line was found for a small sample of
galaxies at z=3 by \citet{Pettini01}.  Even at lower redshifts though
(z$\sim$1), where larger samples and better statistics are available,
the presence of dust is evident and new rather interesting trends
emerge. In an analysis of a large sample of galaxies located at
0.4$\leq$z$\leq$0.8 \citet{Cardiel03} estimated their SFR using the IR
luminosity as well as their H$\alpha$ emission. They found a clear
trend indicating that as one moves from lower to higher luminosity
systems (L$_{\rm IR}=10^{11}$--$10^{12}$\,L$_{\sun}$) the SFR$_{\rm
H\alpha}$ is up to {\em an order of magnitude less} than the SFR$_{\rm
IR}$ (see their Fig. 15) in agreement with the original work by
\citet{Armus90} based on IRAS data.

The result of \citet{Cardiel03} further supports our concern that the
effects of dust in the z$\sim$2 IR luminous systems mentioned earlier
may not have been properly taken into account. It also stresses the
need to better quantify the correlation between the rest frame mid-IR
emission and the L$_{\rm IR}$ of IR luminous galaxies \citep{Chary01},
since a strong active galactic nucleus (AGN) can also contribute to
the mid-IR \citep[see][]{Laurent00,Tran01}.

\subsection{Arp\,299 and VV\,114 analogues in high-z surveys}

To better illustrate another aspect of the influence of dust in the
morphology of an IR luminous galaxy, we decided to compare the UV and
7$\mu$m images of Arp\,299 and VV\,114, the two nearest interacting
pairs of galaxies with L$_{\rm IR}>10^{11}$, which at the same time
are sufficiently extended in the sky (see Fig.~1). Both systems show a
clearly resolved eastern and western component. As they have been
studied extensively in all wavelengths we had access to published high
resolution UV images with HST, mid-IR images with Keck
\citep{Soifer01}, as well as our own deep mid-IR images obtained with
ISOCAM \citep[see][ and references therein]{LeFloch02,Charmandaris02}.

The selection of the rest frame $\sim$0.2\,$\mu$m (UV) and
$\sim$7\,$\mu$m (mid-IR) bands was determined by the fact that at
z$\sim$2 these would be observed in R-/I-bands and at $\sim$24\,$\mu$m
respectively.  The ongoing NOAO Deep Wide-Field Survey\footnote{For
more information on the NOAO Deep Wide-Field Survey visit
http://www.noao.edu/noao/noaodeep} \citep{Jannuzi99} is covering those
optical bands and it is sufficiently sensitive to detect the rest
frame UV emission from Arp\,299 and VV\,114 (see Table 1), as well as
any L$^*$ star-forming galaxy at z$\ge$2 (detection limits of
$\sim$26\,mag in R and I-bands).  A number of deep mid-IR surveys have
also been scheduled to be performed with the Space Infrared Telescope
Facility (SIRTF) using the 24\,$\mu$m imaging using the Multiband
Imaging Photometer (MIPS\footnote{For more information on MIPS and its
sensitivity visit http://mips.as.arizona.edu.})
\citep{Rieke00,Lonsdale03}. SIRTF will readily detect
$\sim$10$^{11}$\,L$_{\sun}$ starburst and star forming galaxies at
z=2, since the rest frame 7.7\,$\mu$m Polycyclic Aromatic Hydrocarbon
(PAH) feature, which is very strong in starburst systems
\citep[see][]{Laurent00}, will move into the MIPS 24\,$\mu$m band.

 At a redshift z=2, though, one arcsecond on the sky, the typical
resolution of ground imagery, corresponds to a projected distance of
8\,kpc. The spatial resolution of SIRTF is even poorer ($\sim$6.9$''$
at 24\,$\mu$m). As a result, systems similar to Arp\,299 and VV\,114
will be unresolved in these surveys.

Observing the two galaxies in Fig. 1 the striking difference between
the UV and 7$\mu$m emission is obvious. In Arp\,299 we note that
$\sim$20\% of the UV emission originates from the eastern component
(IC\,694) and $\sim$80\% is associated with NGC\,3690. However, at 7
and 15\,$\mu$m IC\,694 becomes progressively more luminous relative to
NGC\,3690, contributing $\sim$30\% and $\sim$70\% of the flux
respectively. IC\,694 eventually dominates the IR luminosity of
Arp\,299, contributing twice as much to it than NGC\,3690. As we can
see by the displaced centroid between the mid-IR and UV emission,
discussed in \cite{Charmandaris02}, all UV and optical emission
detected from IC\,694 is probably due to surface star formation
activity. The emission from the massive stars formed by the high
density molecular gas in the nucleus of IC\,694 only becomes visible in
the near-IR and mid-IR wavelengths. The situation in VV\,114 is even
more extreme. Although VV\,114E is 2.5 times brighter than VV\,114W in
the mid-IR it is practically invisible in the UV!  VV\,114E actually
displays strong evidence that a considerable fraction of its mid-IR
flux originates from an AGN \citep{LeFloch02,Soifer01}, which further
enhances its mid-IR emission. The relative contribution of both galaxy
components to the total emission of each system in the the UV and
mid-IR is presented in Table~1.

In Table 1 we also include the 7\,$\mu$m to UV flux density ratio,
$X^{MIR}_{UV}$, for each galaxy, as well as the global value for each
interacting pair. A scatter in those values is substantial. We observe
that the highly extinct galaxy of the pair has an $X^{MIR}_{UV}$ ratio
which can be 1.7 to 24 times greater than the corresponding ratio of
the whole Arp\,299 or VV\,114 respectively!  Furthermore, in both
cases the extreme characteristics of the redder/eastern galaxy, which
is actually responsible for the bulk of the L$_{\rm FIR}$ of the
system, remain inconspicuous if we restrict our observations to the
rest frame UV and optical observations.  When considered as a single
system the UV emission from the less obscured component dilutes the
signature of warm dust emission. Hence, if one is to use the
$X^{MIR}_{UV}$ reddening or even estimate the slope $\beta$ of the UV
spectrum to establish a $\beta$ to L$_{FIR}$ correlation as tracer of
extinction and total dust content \citep[i.e.,][]{Meurer99} in IR
luminous systems, a systematic underestimate of the dust content will
be introduced\footnote{Unfortunately, lack of high spatial resolution
UV and mid-IR maps for a statistically significant sample of local IR
luminous galaxies makes a quantitative prediction of this scatter
impossible.}.

This fact has serious implications when one tries to examine the
colors of distant high redshift sources using deep surveys with low or
moderate spatial resolution. All IR luminous galaxies are very likely
interacting systems. A significant fraction of them, similarly to the
two cases presented above, will remain as unresolved point
sources. Even though the actual value of the rest frame mid-IR to UV
ratio depends on the details of the filters used to observe the high-z
sources (R,I, and/or MIPS24), and they are beyond the scope of this
letter, one result is clear from this exercise. A simple diagnostic
which correlates integrated rest frame UV with mid-IR or radio
measurement is bound to introduce a systematic scatter not just due to
the unknown quantities of dust at high-z, but principally due to the
blending of the emission from the unresolved interacting
components. Depending on the stage of the interaction and the amount
of spatially extended star formation activity, there will be cases
(such as VV\,114) where this scatter may be extreme. The situation
will be even more complicated if an obscured AGN is present since this
will increase the 3--6\,$\mu$m mid-IR emission but it will have
minimal effects in the UV flux. To address these limitations,
follow-up observations with deep near-IR sub-arcsec imaging using the
Hubble Space Telescope or high quality ground facilities will be
necessary in order to reveal whether those systems are indeed
substantially disturbed and help us understand their physical
properties.

\subsection{Frequency of ``VV\,114''-type systems at high-z}

An interesting question to address would be how important/numerous
galaxies with properties similar to VV\,114 are at high-z. The fact
that the extreme characteristics of VV\,114 were realized based on
high resolution UV and mid-IR maps, which are available for only a
handful of the nearby LIRGs/ULIRGs, makes such a prediction
challenging. However, if we use the 1Jy sample of \citet{Kim02} as a
guide, we find that $\sim$10\% of the systems have interacting pairs
with similar spatial separation to VV\,114 and an R-K color of one
galaxy in the system being more than 2 mags different than the other
one. For VV\,114 this color difference is ~3.5\,mag. Even though LIRGs
and ULIRGs are rare in the local Universe contributing to less than
6\% of the local infrared emission \citep[see][]{Sanders96}, they are
much more frequent at high-z. The ISOCAM deep surveys clearly
demonstrate that at z$\sim$1 nearly $\sim$65\% of the peak and
integrated cosmic infrared background originates from galaxies. Of
those galaxies, $\sim$75\% are LIRGs and ULIRGs \citep[see review
by][]{Elbaz03}. Furthermore, as we mentioned earlier, it is accepted
that LIRGs/ULIRGs are the result of galaxy merging and it has been
found that even though interacting systems comprise only 3--9\% of all
local galaxies \citep{Arp75,Struck99}, the merger rate increases
rapidly with redshift (1+z)$^{4-6}$ out to z$\sim$3
\citep{Conselice03} and most high-z galaxies appear somewhat
perturbed.  Based on the above results a simple scaling would suggest
that systems such as VV\,114 may contribute $\sim$5--10\% to the
galaxy counts in the infrared at z$\ge$1. So even though extremely
rare locally, systems such as VV\,114 may have profound effects in our
understanding of the early Universe.

\section{Conclusions}

Using archival images of VV\,114 and Arp\,299, two nearby violently
interacting infrared luminous galaxies, we show the strikingly
different morphology between their 0.2\,$\mu$m UV and 7\,$\mu$m mid-IR
emission. Even though the redshifted rest frame emission from both
wavelengths will be easily detected at z$\sim$2 by ground based
R-/I-band surveys and by the 24\,$\mu$m surveys of SIRTF respectively,
the fact that we will not be able to spatially resolve the different
components of each source will introduce a scatter in their measured
colors. This scatter will not only depend on the relative contribution
of each interacting component to the total IR luminosity, but it will
also cause a systematic underestimate of the integrated rest frame
mid-IR to UV colors leading to lower values of the total dust content
of those systems.

\acknowledgments

We thank the anonymous referee whose comments helped us improve this
paper. VC would like to thank J.R. Houck (Cornell), D. Elbaz
(CEA/Saclay), L. Armus (Caltech), J.Hibbard (NRAO), and J.D. Smith
(Arizona) for useful discussions, as well as to acknowledge the support
of JPL contract 960803.

\newpage 

\clearpage

\begin{deluxetable}{lcccccc}
\tabletypesize{\scriptsize} 
\tablecaption{Rest Frame near-UV and mid-IR emission from Arp\,299 and VV\,114\label{tbl}}
\tablewidth{0pc}
\startdata \\
\tableline 
\tableline \\
			& Arp\,299 & IC\,694 & 	NGC\,3690 & VV\,114 &
			VV\,114E & VV\,114W \\
\tableline \\
UV ($\sim$0.22\,$\mu$m) (mag)\tablenotemark{a}	& 12.73   & 14.55 & 12.95  & 14.49	& 18.38    & 14.53 \\
mid-IR ($\sim$7\,$\mu$m) (mJy)\tablenotemark{b}	& 1032    & 325	  & 707	   & 202 	& 136      & 66 \\
$X^{MIR}_{UV}$=7\,$\mu$m / UV \tablenotemark{c}	        & 196	  & 330	   & 164	& 34	& 819      & 11 \\
\tablenotetext{a}{UV magnitudes based on Table 6 of \citet{Meurer95} and Table 3 of \citet{Goldader02}.}
\tablenotetext{b}{The 7$\mu$m (LW2) ISOCAM flux density over the same regions using data from \citet{Charmandaris02} and \citet{LeFloch02}.}
\tablenotetext{c}{The non dimentional $X^{MIR}_{UV}$ ratio is calculated after converting the UV mag to $f_{\nu}$ in mJy.}
\enddata
\end{deluxetable}

\clearpage 
\begin{figure} 
\figurenum{1} 
\plotone{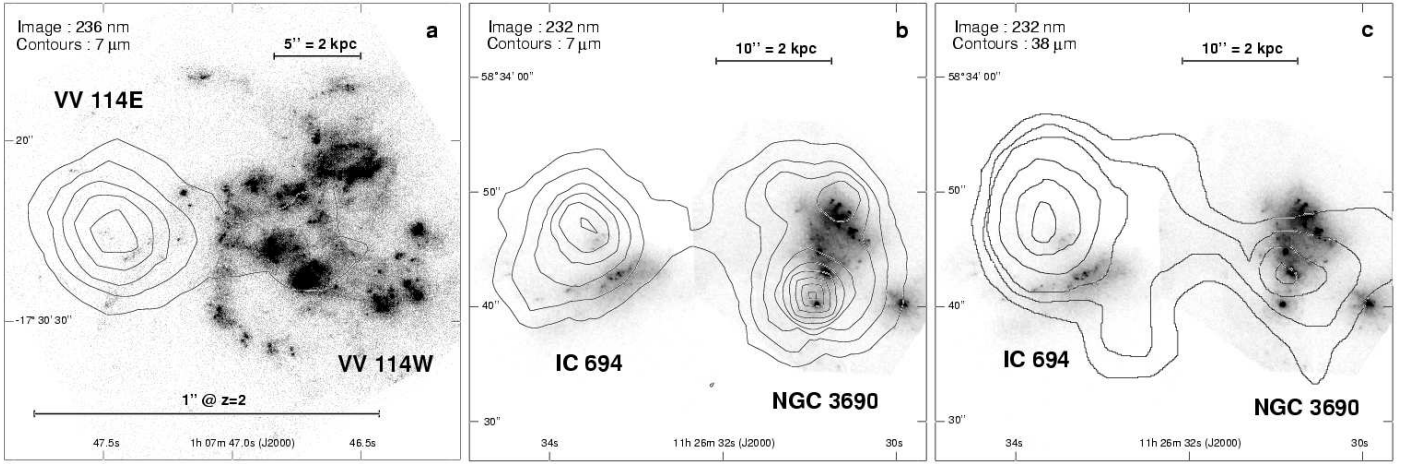}

  \caption{a) An HST/STIS image of VV\,114 at 0.23\,$\mu$m, adapted
  from \citet{Goldader02} with an overlay of the 7$\mu$m emission from
  \citet{LeFloch02}. The images are uniform and the UV flux limit is
  26.6\,mag\,arcsec$^{-2}$. The contour levels are set with
  logarithmic spacing between 0.8 and 8.4\,mJy\,arcsec$^{-2}$ (
  1$\sigma$ $\sim$ 0.15\,mJy\,arcsec$^{-2}$). b) An archival HST/FOC
  (0.22\,$\mu$m) UV image of Arp\,299 with the 7$\mu$m emission
  contours from \citet{Charmandaris02}. The images are uniform and the
  UV flux limit is 21\,mag\,arcsec$^{-2}$.The contour limits are set
  to 1.2 and 30.7\,mJy\,arcsec$^{-2}$ (1$\sigma$ $\sim$
  0.27\,mJy\,arcsec$^{-2}$). Note the offset in the 7$\mu$m centroid
  of IC\,694 when compared to the underying UV emission. c) The same
  UV image of Arp\,299 with an overlay of the 38$\mu$m emission from
  \citet{Charmandaris02}.  The levels are 1.5\,Jy\,beam$^{-1}$
  beginning at 3\,Jy\,beam$^{-1}$ (6\,$\sigma$).  We can easily see
  that in both systems the rest frame UV light from the mid-IR
  dominant source is either completely suppressed (VV\,114) or
  marginally detected (Arp\,299), leading to the high scatter in the
  $X^{MIR}_{UV}$ values of Table 1. For reference we include in a) a
  horizontal bar which indicates the physical scale of 8\,kpc that
  1\,arcsec (the typical angular resolution of ground optical deep
  surveys) would cover if VV\,114 were at z=2. Note that the angular
  resolution of SIRTF at 8 and 24\,$\mu$m is $\sim$2.3 and
  $\sim$6.9\,arcsec respectively. \label{fig1}}

\end{figure}

\end{document}